\documentclass{iau_FM}
\usepackage{graphicx}
\usepackage[]{natbib}
\usepackage{hyperref}
\usepackage[utf8]{inputenc}
\hypersetup{
    colorlinks=true,
    linkcolor=blue,
    filecolor=magenta,      
    urlcolor=blue,
    citecolor={blue}
}

\title[Probing restarting activity in hard X-ray selected GRGs] 
{Probing restarting activity in hard X-ray selected giant radio galaxies} 

\author[G. Bruni, F. Ursini, et al.]   
{G. Bruni$^1$\thanks{website: \href{http://gral.iaps.inaf.it}{http://gral.iaps.inaf.it}},
F. Ursini$^2$, 
F. Panessa$^1$,
L. Bassani$^2$,
A. Bazzano$^1$,
A.~J.~Bird$^3$,
E. Chiaraluce$^1$,
D. Dallacasa$^{4,5}$,
M. Fiocchi$^1$,
M. Giroletti$^5$,
L. Hern\'andez-Garc\'ia$^6$,
A. Malizia$^2$,
M. Molina$^2$,
L. Saripalli$^7$,
P.~Ubertini$^1$,
\and T. Venturi$^5$
 }

\affiliation{
$^1$INAF - Istituto di Astrofisica e Planetologia Spaziali \\ via del Fosso del Cavaliere 100, 00133 Roma, Italy \\ email: {\tt gabriele.bruni@inaf.it} \\[\affilskip]
$^2$INAF - Osservatorio di Astrofisica e Scienza dello Spazio \\ via Piero Gobetti 93/3, 40129 Bologna, Italy \\  email: {\tt francesco.ursini@inaf.it} \\[\affilskip]
$^3$School of Physics and Astronomy, University of Southampton, SO17 1BJ, UK \\[\affilskip]
$^4$DIFA - Dipartimento di Fisica e Astronomia \\ Universit\`a di Bologna, via Gobetti 93/2, 40129 Bologna, Italy \\[\affilskip]
$^5$INAF - Istituto di Radioastronomia, via Piero Gobetti 101, 40129 Bologna, Italy \\[\affilskip]
$^6$IFA - Instituto de F\'isica y Astronom\'ia \\ Universidad de Valpara\'iso, Gran Breta\~na 1111, Playa Ancha, Valpara\'iso, Chile \\[\affilskip]
$^7$Raman Research Institute, C. V. Raman Avenue, Sadashivanagar, Bangalore 560080, India
}

\pubyear{2018}
\jname{Astronomy in Focus, Volume 1} 
\editors{Volker Beckmann \& Claudio Ricci, eds.}
\begin{document}
\maketitle


\begin{abstract}

With their sizes larger than 0.7 Mpc, Giant Radio Galaxies (GRGs) are the largest individual objects in the Universe. To date, the reason why they reach such enormous extensions is still unclear. One of the proposed scenarios suggests that they are the result of multiple episodes of jet activity. Cross-correlating the INTEGRAL+Swift AGN population with radio catalogues (NVSS, FIRST, SUMSS), we found that 22\% of the sources are GRG (a factor four higher than those selected from radio catalogues). Remarkably, 80\% of the sample shows signs of restarting radio activity. The X-ray properties are consistent with this scenario, the sources being in a high-accretion, high-luminosity state with respect to the previous activity responsible for the radio lobes.

\end{abstract}


\firstsection 
              
\section{Introduction}

A relatively small fraction of powerful radio galaxies ($\sim$6\% in the 3CR catalogue, \citealt{Ishwara}) exhibits rather large linear extents, i.e. above 0.7 Mpc, making them the largest individual objects in the Universe. These sources are usually referred to as Giant Radio Galaxies (GRG), and can exhibit both Fanaroff-Riley type I and type II radio galaxies (FRI and FRII respectively, \citealt{Fanaroff}). While FRI GRGs are associated with early type galaxies, those with FRII morphology are hosted both in early type galaxies and quasars. 
The samples of GRGs available in the literature, mainly drawn from all sky radio surveys such as NVSS, SUMSS, WENSS, have been used to test models for radio galaxy evolution and investigate the origin of such incredibly extended structures (i.e. \citealt{Blundell}). Despite the dynamical ages typically overestimate the radiative age of the radio source by a factor 2-4 (see Fig 5 in \citealt{Parma}), the general correlation observed between size and age in radio galaxies (Fig 6 in \citealt{Parma}) suggests that GRGs represent the oldest tail of the age distribution for radio galaxies.
Beyond the source age, the main intrinsic parameters that allow a radio galaxy to reach a linear size of the order of a Mpc during its lifetime are still unclear. The medium must play a role in the overall jet expansion, but its effects remains difficult to evaluate, not to mention that the density of the medium explored by the radio jet during its life/development may change considerably over the large scales considered here. Some GRGs are associated with the dominant member of a galaxy group (e.g. the FRI-GRG NGC 315, \citealt{Giacintucci}), while others have been detected at high redshift in a likely less dense environment (\citealt{Machalski}). Those authors also concluded that the jet power and the central density of the galaxy nucleus seem to correlate with the size of radio galaxies. Yet another study, based on optical spectroscopy of galaxies in a large-scale environment around the hosts of 19 GRGs (\citealt{Malarecki}) finds a tendency for their lobes to grow to giant sizes in directions that avoid dense galaxy on both small and large scales. Finally, other authors suggested that GRG could reach their size thanks to more than one activity episode, being restarting radio sources (\citealt{Sub}). More recently, progress in the study of this class of sources has been achieved thanks to the use of low frequency facilities - such as LOFAR, JVLA and GMRT - where old relativistic plasma is better seen (e.g. \citealt{Orru, Clarke, Sebastian}). All in all, however, the origin and evolution of GRGs remains until now very much unconstrained.


\section{Hints of restarting activity}

Starting from 2002, the hard X-ray and $\gamma$-ray sky has been surveyed by \emph{INTEGRAL}/IBIS and \emph{Swift}/BAT in the spectral range from 10-200 keV. Up to now many catalogues have been released, the most recent ones comprising more than 1000 high energy sources (\citealt{Bird,Oh}), with a large fraction of objects unambiguously associated with AGN. Our group is carrying out a multi-wavelength study of a sample of hard X-ray selected GRG extracted from these high energy catalogues.
\cite{Bassani} undertook a radio/$\gamma$-ray study of the combined \emph{INTEGRAL}+\emph{Swift} AGN populations, and found 64 sources associated with extended radio galaxies with measured redshift. They belong to both the FRI and FRII morphological classes.
Interestingly, inspection of NVSS and SUMSS revealed that 14 of them are GRG, i.e. $\sim$22\% of the sample. Considered the classical fraction of giant sources in radio-selected samples of radio galaxies (1-6\%), this fraction is impressive, and suggests a tight link between the nuclear/accretion properties of the AGN and the radio source size.

In order to better characterize the GRG in this sample and to study their evolutionary history, we are collecting multi-band high sensitivity observations for the fraction of sample observable from the Northern hemisphere (12 targets). Remarkably, 10 over 12 sources ($\sim$80\%) of the sample show signs of restarting radio activity from previous studies in the literature or our radio campaign (\citealt{Bruni}), with one even showing an extreme reorientation from radio-galaxy to BL Lac (\citealt{Hernandez}). This large fraction suggests that multiple radio phases could justify the large size, and be responsible for the strong hard X-ray emission coming from a possible refueled radio core - as suggested from X-ray data (\citealt{Ursini}). Our aim is to confirm the restarting activity for these sources, and possibly understand whether this scenario is a distinctive property of hard X-ray selected GRGs, or rather a general property of GRGs. In 2014, we have collected GMRT deep observations at 325 and 610 MHz for a pilot sample of 4 sources from our GRG sample. Those GMRT data allowed our group to identify the second known giant X-shaped radio galaxy (IGR J14488-4008: \citealt{Molina15}), and a newly discovered GRG (IGR J17488-2338: \citealt{Molina14}).



\begin{figure}
    \includegraphics[width=0.5\textwidth]{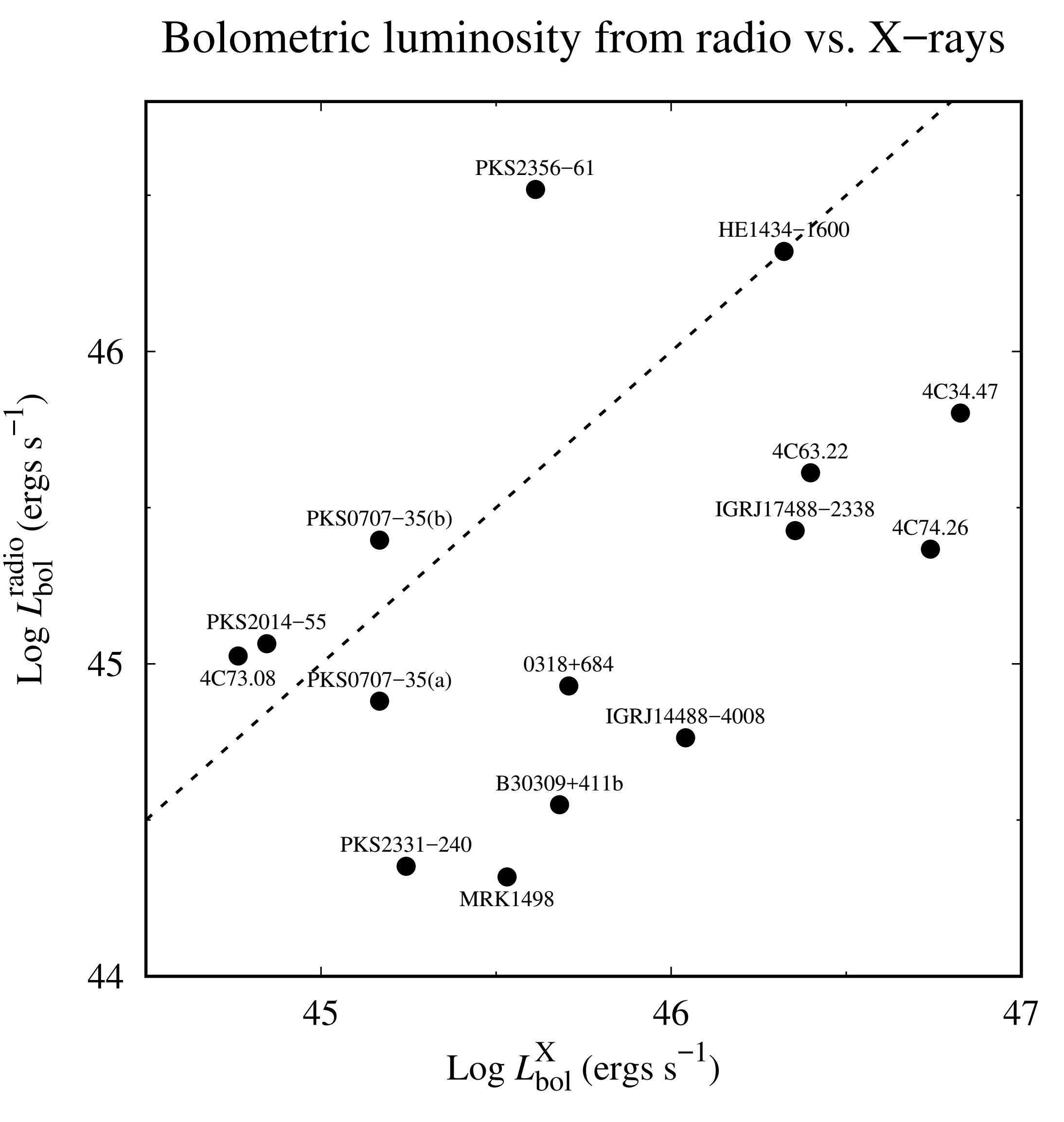}
    \includegraphics[width=0.5\textwidth]{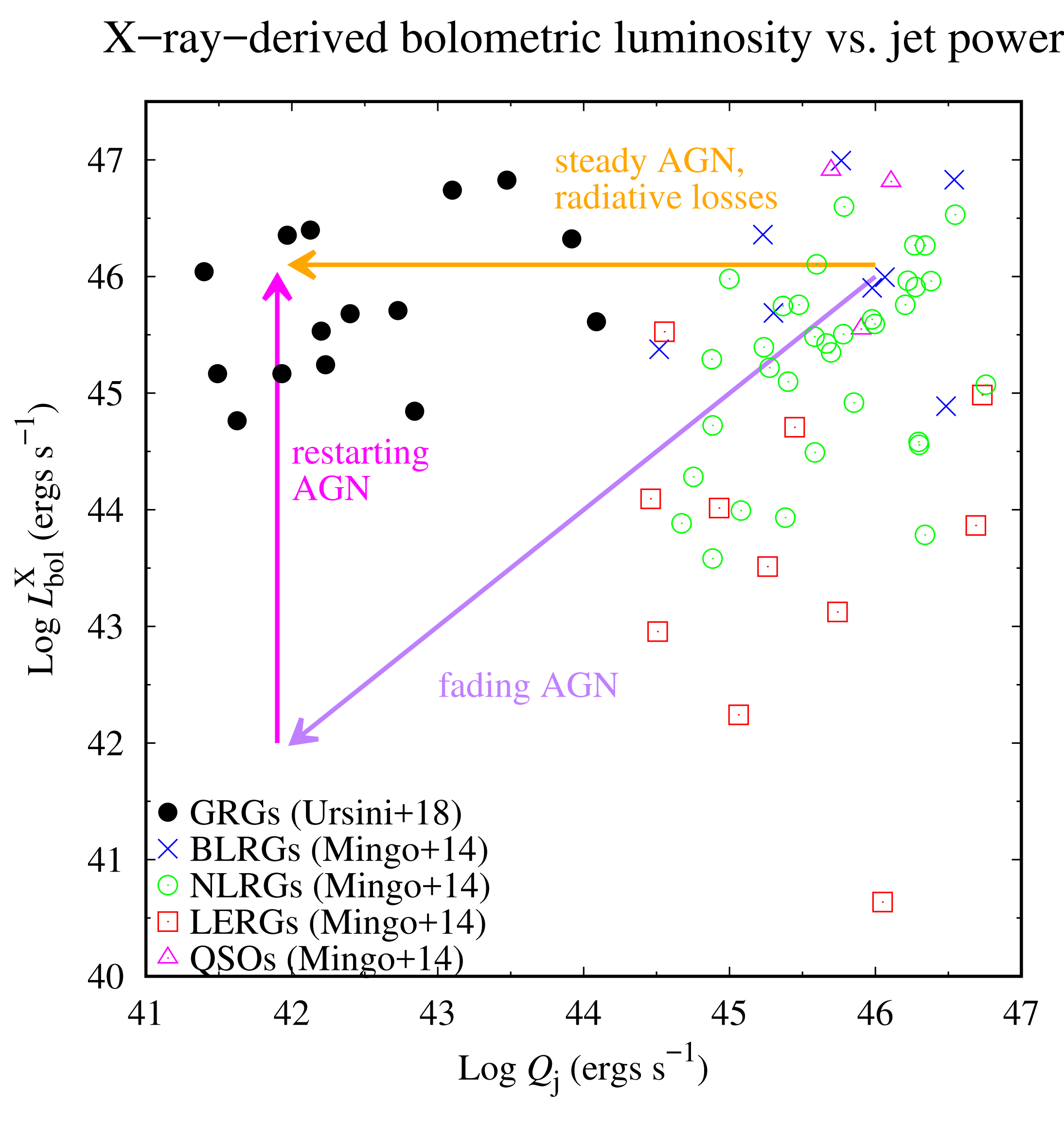}
    \caption{ \textit{Left panel:} Bolometric luminosity estimated from the radio luminosity of the
lobes versus that estimated from the 2--10 keV luminosity.
The dashed line represents the identity $y = x$. \textit{Right panel:} Bolometric luminosity estimated from the 2--10 keV luminosity versus jet power estimated from the relation of Willott et al. (1999). Black dots denote the GRGs of our sample, overlayed in the plot of Mingo et al. (2014). The colored arrows represent two putative evolutionary paths of radio galaxies. Both panels are adapted from \cite{Ursini}.}
    \label{fig:lobes-x}
\end{figure}

\section{X-ray properties}

The bulk of the X-ray emission of GRGs in our sample is consistent with originating from a Comptonizing corona coupled to a radiatively efficient accretion flow (Eddington ratio $>$ 0.02) (\citealt{Ursini}), like in normal-size FR II radio galaxies. This indicates that the nuclei are currently active, despite the likely old age of the radio lobes.
The peculiar morphology makes it possible to study the relation between the X-ray emission and the radio emission in detail, separating the contribution from the core and from the lobes. We find that:
\begin{itemize}
    \item[-] The X-ray luminosity $L_{\textrm{\scriptsize 2--10keV}}$ correlates with the radio core luminosity $L_{\textrm{\scriptsize 1.4GHz}}^{\textrm{\scriptsize core}}$, as expected from the so-called fundamental plane of black hole activity (\citealt{Merloni}). The slope of the correlation is consistent with the `radiatively efficient' branch of the fundamental plane rather than the `standard/inefficient' branch (\citealt{Coriat}).
    
    \item[-] In most sources, the X-ray luminosity yields an estimate of the bolometric luminosity ($L_{\textrm{\scriptsize bol}}^{\textrm{\scriptsize X}}$) an order of magnitude larger than the corresponding estimate from the radio lobes luminosity ($L_{\textrm{\scriptsize bol}}^{\textrm{\scriptsize radio}}$), from the relation of \cite{vv} (Fig. \ref{fig:lobes-x}, left panel).
    
    \item[-] The time-averaged kinetic power of the jets, as estimated from the radio luminosity using the relation of \cite{Willott}, is much lower than in the radio luminous AGNs studied by \cite{Mingo}. This discrepancy is up to 3 orders of magnitude, while the bolometric luminosity is perfectly consistent with high-excitation radio galaxies (Fig. \ref{fig:lobes-x}, right panel).
\end{itemize}
These results are consistent with a restarting activity scenario, i.e. the sources are currently highly accreting and in a high-luminosity state compared with the past activity that produced the old and extended radio lobes. GRGs could start their life with high nuclear luminosities and high jet powers; with time, the central engine gradually fades while the radio lobes expand. Eventually, the nuclear activity can be triggered again following a new accretion episode, resulting in a strong increase of the core luminosity. 
Alternatively, the nuclei would need to sustain a steady activity during their lifetime, with a nearly constant accretion rate and core luminosity; radiative losses would produce the observed dimming of the radio lobes as they grow in size and interact with the environment. In this case, however, the nuclei are required to stay active for at least 100--250 Myrs (\citealt{Machalski}).


\section{Future work}
Triggered by these results, we are collecting more data both in radio and X-ray bands, in order to connect the nuclear accretion status to the Mpc-scale structure of these objects. The cores have been observed in single-dish mode (Effelsberg-100m telescope) in June 2018, with the aim of reconstructing the radio SED and test the fraction of young radio components. In the X-ray band, a \emph{Swift}/XRT campaign has been planned to build a complete comparison sample of radio-selected GRG \citep{schoen}, and highlight the different properties with respect to the hard X-ray selected ones. Indeed, radio selected GRGs are not necessarily X-ray bright, thus allowing us to explore new portions of the parameter space in luminosity-luminosity diagrams such as those in Fig. \ref{fig:lobes-x}.



\begin{thebibliography}{}

\bibitem[Bassani et al. (2016)]{Bassani}{Bassani, L., Venturi, T., Molina, A. et al.} 2016, \textit{MNRAS}, 461, 3165

\bibitem[Bird et al. (2016)]{Bird}{Bird, A. J., Bazzano, A., Malizia, A. et al.} 2016, \textit{MNRAS}, \textit{ApJS}, 223, 15

\bibitem[Blundell et al. (1999)]{Blundell}{Blundell, K. M., Rawlings, S., \& Willott, C. J.} 1999, \textit{AJ}, 117, 677

\bibitem[Bruni et al. (2018)]{Bruni}{Bruni, G., Panessa, F., Bassani, G. et al.} 2018, \textit{MNRAS}, in prep.

\bibitem[Clarke et al. (2017)]{Clarke}{Clarke, A.O., Heald, G., Jarrett, T. et al.} 2017, \textit{A\&A}, 601, A25

\bibitem[Coriat et al. (2011)]{Coriat}{Coriat M., Corbel, S., Prat, L. et al.} 2011, \textit{MNRAS}, 414, 677

\bibitem[Fanaroff \& Riley (1974)]{Fanaroff}{Fanaroff, B. L. \& Riley, J. M.} 1974, \textit{MNRAS}, 167, 31

\bibitem[Giacintucci et al. (2011)]{Giacintucci}{Giacintucci, S., O'Sullivan, E., Vrtilek, J. et al.} 2011, \textit{ApJ}, 732, 95

\bibitem[Hern\'andez-Garc\'ia et al. (2017)]{Hernandez}{Hern\'andez-Garc\'ia, L., Panessa, F., Giroletti, M. et al.} 2017, \textit{A\&A}, 603, A131

\bibitem[Ishwara-Chandra \& Saikia (1999)]{Ishwara}{Ishwara-Chandra, C.H. \& Saikia, D.J.} 1999, \textit{MNRAS}, 309, 100

\bibitem[Machalski et al. (2004)]{Machalski}{Machalski, J., Chyzy, K. T. \& Jamrozy, M.} 2004, \textit{Acta Astronomica}, 54, 249

\bibitem[Malarecki et al. (2004)]{Malarecki}{Malarecki, J. M., Jones, D. H., Saripalli, L. et al.} 2015, \textit{MNRAS}, 449, 955

\bibitem[Merloni et al. (2003)]{Merloni}{Merloni A., Heinz S., di Matteo T. et al.} 2003, \textit{MNRAS}, 345, 1057

\bibitem[Mingo et al. (2014)]{Mingo}{Mingo, B., Hardcastle, M. J., Croston, J. H. et al.} 2014, \textit{MNRAS}, 440, 269

\bibitem[Molina et al. (2014)]{Molina14}{Molina, M., Bassani, L., Malizia, A. et al.} 2014, \textit{A\&A}, 565, A2

\bibitem[Molina et al. (2015)]{Molina15}{Molina, M., Venturi, T., Malizia, A. et al.} 2015, \textit{MNRAS}, 451, 3

\bibitem[Oh et al. (2018)]{Oh}{Oh, K., Koss, M., Markwardt, C.B. et al.} 2018 \textit{ApJS}, 234, 4

\bibitem[Orr\`u et al. (2010)]{Orru}{Orr\`u, E., Murgia, M., Feretti, L., et al.} 2010, \textit{A\&A}, 515, A50

\bibitem[Parma et al. (1999)]{Parma}{Parma, P., Murgia, M., Morganti, R. et al.} 1999, \textit{A\&A}, 344, 7

\bibitem[Schoenmakers et al.(2000)]{schoen}{Schoenmakers, A.~P., Mack, K.-H., de Bruyn, A. G., et al.} 2000, \textit{A\&AS}, 146, 293 

\bibitem[Sebastian et al. (2018)]{Sebastian}{Sebastian, B., Ishwara-Chandra, C. H., Joshi, R. et al.} 2018, \textit{MNRAS}, 473, 4

\bibitem[Subrahmanyan et al. (1996)]{Sub}{Subrahmanyan, R., Saripalli, L., Hunstead, R.W.} 1996, \textit{MNRAS} 279, 257

\bibitem[Ursini et al. (2018)]{Ursini}{Ursini, F., Bassani, L., Panessa, F. et al.} 2018, \textit{MNRAS}, in press

\bibitem[van Velzen et al. (2015)]{vv}{van Velzen S., Falcke H., K\"ording E. et al.} 2015, \textit{MNRAS}, 446, 2985

\bibitem[Willot et al. (1999)]{Willott}{Willott, C. J., Rawlings, S., Blundell, K. M. et al. 1999}, \textit{MNRAS}, 309, 1017

\end{thebibliography}
\end{document}